\documentclass[final]{aipproc}
\layoutstyle{8x11double}
\usepackage{amsmath}
\begin{document}

\title{Logarithmic decay in a two-component model}
\author{Matthias Sperl}{address={Physik-Department, 
Technische Universit\"at M\"unchen, 85747 Garching, Germany}}

\begin{abstract}
The correlation functions near higher-order glass-transition singularities are 
discussed for a schematic two-component model within the mode-coupling theory 
for ideal 
glass-transitions. The correlators decay in leading order like $-\ln(t/\tau)$ 
and the leading correction introduces characteristic convex and concave 
patterns in the decay curves. The time scale $\tau$ follows a Vogel-Fulcher 
type law close to the higher-order singularities.
\end{abstract}

\maketitle

\section{Introduction}

Mode-coupling theory for ideal glass-transitions (MCT) describes the 
transition from a liquid to a glass as a bifurcation in the equation for 
the long-time limit of the density-autocorrelation function 
$\phi_q(t)=\langle\rho_q^*(t)\rho_q\rangle/\langle|\rho_q|^2\rangle$ for
density fluctuations of wave-number modulus $q=|\vec{q}|$. If the 
long-time limit $f_q=\lim_{t\rightarrow\infty}\phi_q(t)$ reaches zero, 
$f_q=0$, the system is in the liquid state. For $f_q>0$, the system is 
in a glass state \cite{Bengtzelius1984}. The theory has been worked out in
detail for the hard-sphere system (HSS) \cite{Bengtzelius1984,Franosch1997}
where the results of experiments \cite{Megen1995} and computer simulations
\cite{Kob2003} support the validity of MCT.
In addition to liquid-glass transitions, MCT also allows for bifurcations of 
higher order \cite{Goetze1988b}. It can be shown that only bifurcations of
a certain hierarchy $A_\ell$ can arise from MCT \cite{Goetze1995b} which are 
equivalent to the singularities in the real roots of polynomials of order 
$\ell$ upon variation of parameters \cite{Arnold1986}. The simplest singularity
$A_2$ or \textit{fold} generically occurs when only a single control parameter 
is changed and the liquid-glass transition is identified with this bifurcation.
At the critical value of the control parameter the long-time limit $f_q$ jumps
from zero to $f_q^c$. Close to the singularity and for small $|\phi_q(t)-f_q|$,
the correlation functions can be expanded in asymptotic series yielding
power laws. A detailed analysis including leading and next-to-leading order
results was given for the HSS \cite{Franosch1997,Fuchs1998,Goetze1999b}.

An $A_3$-singularity or \textit{cusp} can appear when two control parameters
are varied. The $A_3$-singularity is the endpoint of a line of 
$A_2$-singularities which are identified as glass-glass-transition points where
a first glass state, characterized by $f^1_q>0$, transforms discontinuously 
into a second glass state with $f_q^c>f^1_q$. At the $A_3$, this discontinuity 
vanishes and the glass-glass-transition line ends. 
Such cusp was predicted for particles interacting with a hard core 
and a short-ranged attraction \cite{Fabbian1999,Bergenholtz1999} where density
and strength of the attraction are the control parameters. The 
glass-glass transition takes place between a state dominated by repulsion as
in the HSS and a state dominated by attraction, the latter arrested state
was proposed to be related to a gel \cite{Bergenholtz1999}.

Tuning a third control-parameter, the extension of the glass-glass-transition 
line can be varied and once the latter vanishes, the $A_3$-singularity merges 
with the liquid-glass-transition line and gives birth to a \textit{swallowtail}
or $A_4$-singularity. If the range of the attraction is of the order of 5\% of
the hard-core diameter such singularity is predicted for a square-well system
with a strength of the attraction of several $k_\text{B}T$ and a density which
is slightly larger than the one for the glassy arrest in the HSS 
\cite{Dawson2001}.

In contrast to the power-laws at the $A_2$-singularity, the dynamics
close to higher-order singularities is ruled by the logarithm of the time 
\cite{Goetze1988b}. The wave-vector dependent asymptotic solutions along with
the corrections have been calculated in full generality \cite{Goetze2002}. The 
asymptotic laws were demonstrated for the $A_3$-singularity of a two-component 
model \cite{Goetze2002}, and for the higher-order singularities of microscopic 
models with short-ranged attraction \cite{Sperl2003a}. Logarithmic decays 
compatible with the predicted scenarios were found in recent computer 
simulation studies \cite{Puertas2002,Sciortino2003pre}. 

In the vicinity of an $A_3$-singularity, different liquid-glass-transition 
lines cross, and the dynamics in the liquid regime close to this crossing is 
influenced by three different singularities. Dynamical scenarios predicted by 
theory in that region \cite{Fabbian1999,Dawson2001,Sperl2003bpre} were found in
experiments \cite{Mallamace2000,Pham2002,Chen2002,Poon2003,Chen2003b,Pham2003} 
and computer-simulation studies \cite{Pham2002,Zaccarelli2002b}. These findings
indicate that such rich dynamics is indeed relevant for these colloidal systems
and that asymptotic expansions at the MCT-singularities can interpret the 
scenario qualitatively and explain the data even quantitatively 
\cite{Sperl2003bpre}.

In the following, the two-component schematic model used already in 
Ref.~\cite{Goetze2002} shall be reconsidered with respect to the 
$A_4$-singularity and the evolution of the time scales. The advantage of the 
schematic model is that all transition points can be calculated analytically. 
Even in this simple model typical features of the full microscopic theory are 
identified. First, the asymptotic laws for the higher-order singularities and
the schematic model shall be introduced briefly, further details are found in
Ref.~\cite{Goetze2002}. Second, the asymptotic approximation is shown to work 
well for the correlation functions at the $A_4$-singularity. Third, the time
scales for the logarithmic decays at $A_3$- and $A_4$-singularity are discussed
for specific paths in control-parameter space.

\section{Asymptotic laws}

The $A_2$-singularities are characterized by a two-step relaxation around a 
plateau value, and the asymptotic decay laws are given by scaling functions 
\cite{Franosch1997}. For short times the decay onto the plateau is given by
a critical decay $\phi_q(t)-f_q(t)=h_q(t_0/t)^a$, while for long times one
gets the von~Schweidler law $\phi_q(t)-f_q(t)=-h_q(t/t_\sigma')^b$. The 
exponents are determined by a single exponent parameter 
$\lambda=\Gamma(1-a)^2/\Gamma(1-2a)=\Gamma(1+b)^2/\Gamma(1+2b)<1$. For 
$\lambda=1$, the asymptotic solution by power laws becomes invalid and a
higher-order singularity occurs.
The dynamics at higher-order singularities $A_l,\,l>2$ is described up to 
next-to-leading order by \cite{Goetze2002}
\begin{equation}\label{eq:G1G2q}\begin{split}
\phi_q (t) = (f_q^c + \hat{f}_q) + h_q  \bigl [  (- B + B_1)
\ln (t / \tau)
\\  +  (B_2 + K_q B^2) \ln^2 (t / \tau)
\\ \qquad +     B_3 \ln^3 (t / \tau) +  B_4 \ln^4 (t / \tau)  \bigr ] \,\, .
\end{split}\end{equation}
The wave-vector-dependent numbers $f_q^c$, $h_q$, and $K_q$ are characteristic 
for a specific singularity, while $B$, $B_i$, and $\hat{f}_q$ are in addition 
functions of the separation parameters $\varepsilon_1$ and $\varepsilon_2$
quantifying the distance of the actual state from the higher-order singularity
\cite{Goetze2002}. The leading order in Eq.~(\ref{eq:G1G2q}) is given by the 
first line with $B_1=\hat{f}_q=0$ and the leading order prefactor for the 
logarithm is $B\propto\sqrt{|\varepsilon_1|}$. The time scale $\tau$ is fixed 
by matching the asymptotic approximation with the numerical solution for 
$\phi_q(t)$ at the plateau $(f_q^c + \hat{f}_q)$.

\section{The model}

To mimic the $q$-dependence, which is an important feature of the asymptotic 
expansions above, we use a two-component model that was introduced for the 
description of a symmetric molten salt \cite{Bosse1987b}. The model has 
three control parameters which are combined to the vector 
$\mathbf{V}=(v_1, v_2, v_3)$. We will use Brownian dynamics, so the 
equations of motion for the correlators $\phi_q(t),\;q=1,\,2$, read
\begin{subequations}
\label{eq:BK_eom}
\begin{eqnarray}\label{eq:BK_eom_int}
\tau_q \partial_t \phi_q(t) + \phi_q (t) +
\int_0^t m_q (t - t^\prime) \partial_{t^\prime} \phi_q
(t^\prime)dt^\prime = 0\,,\\
\label{eq:BK_eom_m1}
m_1(t) = v_1 \phi_1^2(t) + v_2 \phi_2^2(t),&&\\
\label{eq:BK_eom_m2}
m_2(t) = v_3 \phi_1(t)\phi_2(t)\,.&&
\end{eqnarray}
\end{subequations}  
For the long-time limit of Eq~(\ref{eq:BK_eom}),
one gets a parameterized representation of the transition surface
\cite{Goetze2002},
\begin{subequations}\label{eq:BK_PD}
\begin{equation}\label{eq:BK_PD_xy}
v_3^{c} = x\,,\quad f_1^{c} = y\,.
\end{equation}
\begin{eqnarray}\label{eq:BK_PD_v1}
v_1^{c} &=& \frac{3 - (2+x) y}{2 (1-y)^2 y (2-x y)}
\,,\\\label{eq:BK_PD_v2}
v_2^{c} &=&  \frac{x^2 y (y^2 - 2 y^3)}{2 (1-y)^2 (x^2 y^2 -3 x y + 2)}  
\,\,.
\end{eqnarray}
The variables $x$ and $y$ with $x>4$ and $1/2 \leq y \leq 3/(2+x)$ 
serve as surface parameters. The exponent parameter $\lambda=1-\mu_2$ is 
determined by
\begin{equation}\label{eq:BK_PD_mu2}
\mu_2 =
\frac{(3 x^2 + 6 x)y^3 - (x^2+18x+8)y^2 + (6x+18)y-6}
                {(2x^2+4x)y^3 - 12 x y^2 + (2x+4)y }
\,\,.
\end{equation} 
\end{subequations}
\begin{figure}[htb]
\includegraphics[width=\columnwidth]{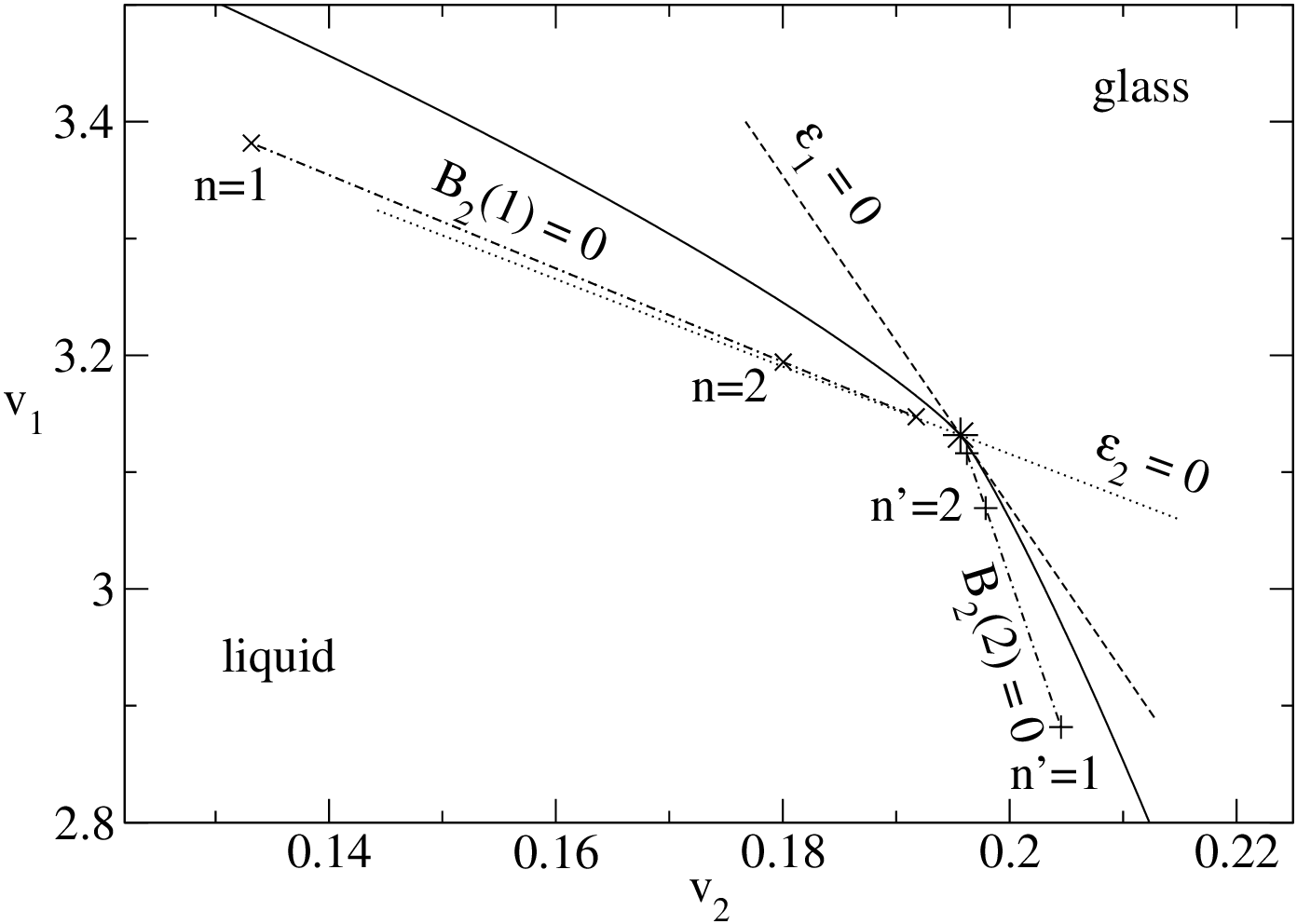}
\caption{\label{fig:BKA4gtd}Glass-transition diagram for the two-component 
model for the cut $v_3=v_3^*$. Liquid-glass transition points are shown by
the full line, the $A_4$-singularity is shown by a star. The vanishing
separation parameters $\varepsilon_1$ and $\varepsilon_2$ are shown by dashed
and dotted straight lines, Chain lines mark the location of vanishing 
correction $B_2(q)=B_2+K_qB^2$ for $q=1,\,2$ with paths indicated by crosses
and pluses, respectively.
}
\end{figure}

The $A_3$-singularities are given by $\lambda=1$ or equivalently $\mu_2=0$
in Eq.~(\ref{eq:BK_PD_mu2}). For $x$ large enough, such a solution exists
with two roots, $y_1(x)<y_2(x)$, where only $y_2(x)$ is relevant.
For small $x$, no such solution exists. Varying $x$, the two cusp values 
$y_1(x)$ and $y_2(x)$ coalesce at $x^*$ with $y_1(x^*)=y_2(x^*)=y^*$, defining
the $A_4$-singularity, $x^*=24.779392\dots,\,y^*=0.24266325\dots$~.

The cut through the transition surface for $v_3 = x^*$ is shown in 
Fig.~\ref{fig:BKA4gtd} as pair of light full lines joining at the 
$A_4$-singularity which is indicated by a star, $(v_1^*, v_2^*, v_3^*) = 
(3.132, 0.1957, 24.78)$. Attached to the $A_4$-singularity we find the lines 
of vanishing separation parameters, $\varepsilon_1(\mathbf{V})=0$ (dashed) and 
$\varepsilon_2(\mathbf{V})=0$ (dotted), which represent a local coordinate 
system. The correction amplitudes at the $A_4$-singularity are $K_1=0.3244$ and
$K_2=-2.109$; these yield two lines of vanishing quadratic correction shown by 
the chain lines in Fig.~\ref{fig:BKA4gtd}.

\begin{figure}[htb]
\includegraphics[width=\columnwidth]{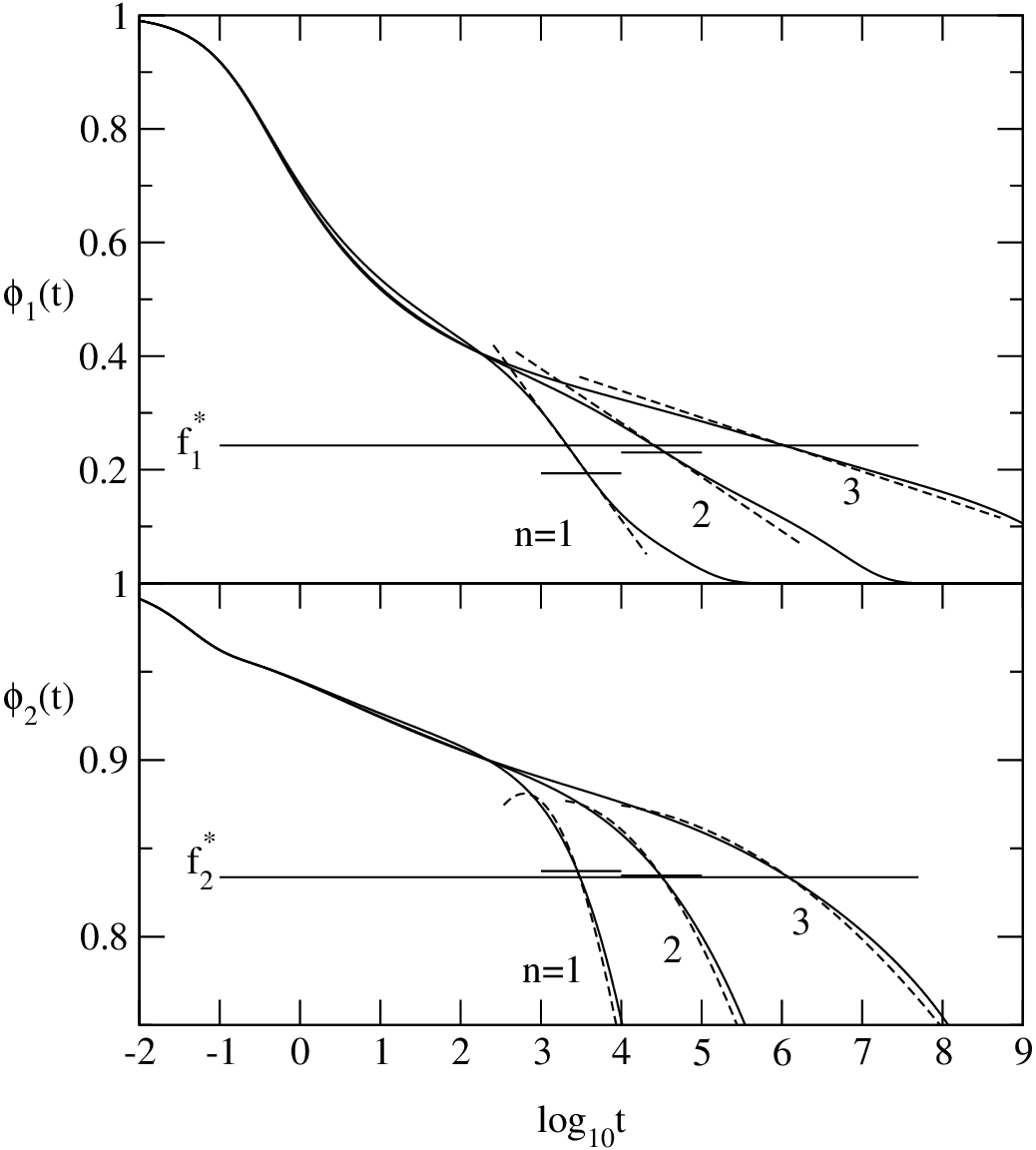}
\caption{\label{fig:A4f1path}Logarithmic decay at the $A_4$-singularity on 
the path ($\times$) with $n=1,\,2,\,3$ in Fig.~\ref{fig:BKA4gtd} with
$v_1-v_1^*=1/2^{n+1}$, $v_2^*-v_2=0.25009/2^{n+1}$. The solutions
of Eq.~(\ref{eq:BK_eom}) are shown as full lines, the 
approximation~(\ref{eq:G1G2q}) as dashed lines. Long horizontal lines
exhibit the critical plateau values $f_q^*$, short lines the corrected
plateau values $f_q^*+\hat{f}_q$.
}
\end{figure}

\section{Asymptotic approximations}

We first analyze the path labeled $n$ ($\times$) in Fig.~\ref{fig:BKA4gtd}. 
The solutions of Eq.~(\ref{eq:BK_eom}) are shown as full curves in 
Fig.~\ref{fig:A4f1path}. The approximations~(\ref{eq:G1G2q}) 
for the correlators are displayed as dashed lines. The time scale $\tau$ 
was matched for $\phi_1(t)$ and $\phi_2(t)$ independently at the 
corrected plateau values. The approximation describes the decay around the 
plateaus $f_q^*$ reasonably for both correlators. Since $B_2+K_1B^2$ is set to 
zero on the present path and $B_3=B_4=0$ for an $A_4$-singularity 
\cite{Goetze2002}, only the first line in Eq.~(\ref{eq:G1G2q}) is relevant for
the upper panel. This approximation describes the decay of $\phi_1(t)$ from 
$0.4$ down to $0.1$ and for a window increasing window in time with increasing 
$n$. Since $K_2<K_1$, the 
prefactor for the quadratic correction in Eq.~(\ref{eq:G1G2q}) for $\phi_2(t)$
is negative and hence the decay around the plateau $f_2^*$ is concave which
is described well by the asymptotic approximation. However, for both $q$ the
prefactor $(B-B_1)$ slightly overestimates the absolute slope of the solutions.
This is due to a relatively large positive next-to-leading-order correction 
$C_1$ that renormalizes the prefactor to $(B-B_1-C_1)$ \cite{Sperl2003}.

Another feature of the curves in Fig.~\ref{fig:A4f1path} is noteworthy. There
appears a window in time outside the transient dynamics between $t\approx 1$ 
and $t\approx 10^3$ where the  description by Eq.~(\ref{eq:G1G2q}) is not 
applicable. This is caused by the close-by $A_2$-singularities on the chosen
path, cf. Fig.~\ref{fig:BKA4gtd}, where $f_q^c>f_q^*,\,q=1,\,2$ which 
introduces an additional slowing down of the dynamics before the logarithmic 
decay is encountered. In addition, we observe that $\phi_2(t)$ varies almost 
linearly in $\ln t$ between $t\approx1$ and $t\approx 250$. It is, however, 
easy to distinguish that effective logarithmic variation, that stays the same 
upon further approaching the singularity, from the characteristic behavior of 
the correlators around the plateau where the prefactor of the logarithmic decay
vanishes with the square root of the distance. The almost linear relaxation in
$\log t$ seen for $\phi_2(t)$ is related to the $\beta$-peak phenomenon
\cite{Goetze2003pre}.

\begin{figure}[htb]
\includegraphics[width=\columnwidth]{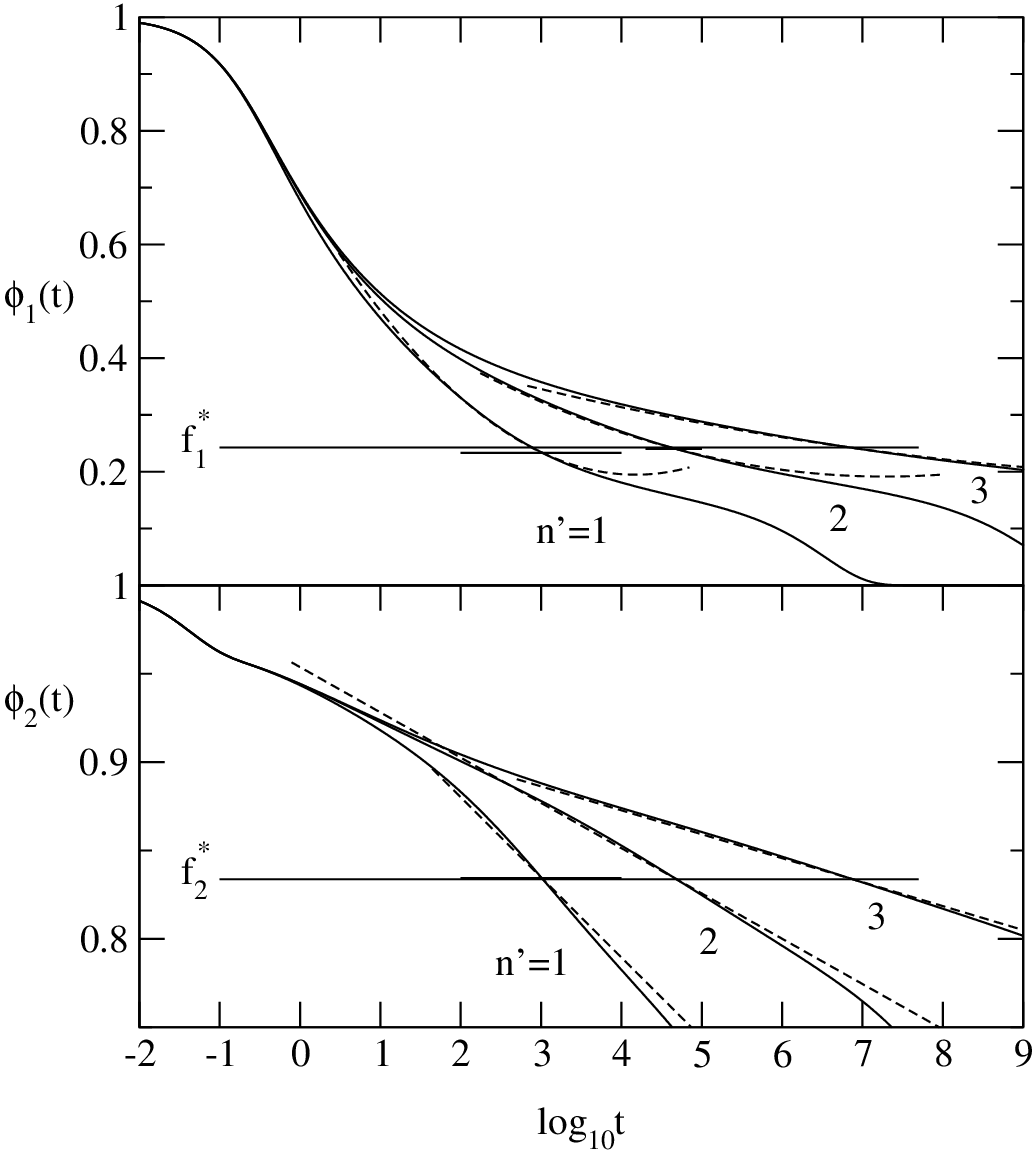}
\caption
{\label{fig:A4f2path}Logarithmic decay on the path ($+$) for 
$n'=1,\,2,\,3$ in Fig.~\ref{fig:BKA4gtd}. $v_1^* -v_1=1/4^n$, 
$v_2-v_2^* = 0.3541/4^n$. Line styles and notation are the same as in 
Fig.~\ref{fig:A4f1path}.
}
\end{figure}
In Fig.~\ref{fig:A4f2path} the quadratic corrections in Eq.~(\ref{eq:G1G2q}) 
for the correlator $\phi_2(t)$ are zero up to higher orders. The quadratic 
corrections to the first correlator are positive, $B_2(1)>0$, and $\phi_1(t)$ 
is convex in $\ln t$ around $f_1^*$. There are close-by $A_2$-singularities 
on the chosen path with $f_q^c<f_q^*,\,q=1,\,2$, causing a decay which is 
encountered after the logarithmic decay, which is seen best for curves $n'=1$
and $n'=2$ in the upper panel of Fig.~\ref{fig:A4f2path} for $\phi_1(t)$<0.15. 
While in Fig.~\ref{fig:A4f1path} the validity of the 
asymptotic approximation around $f_q^*$ was limited by a higher plateau from 
above, in Fig.~\ref{fig:A4f2path} the lower plateaus present a boundary for
the application of Eq.~(\ref{eq:G1G2q}) from below. Because higher plateaus
do not interfere, the asymptotic description applies three decades earlier
in Fig.~\ref{fig:A4f2path} than in Fig.~\ref{fig:A4f1path} for comparable
distances from the $A_4$-singularity.

The estimate for the sign of $C_1$ given above mainly depends on the relative 
distance of the path chosen in control-parameter space to the lines 
$\varepsilon_1=0$ and $\varepsilon_2=0$.
As the ordering of the latter two lines and the path is reversed for the case
$B_2(2)=0$ in Fig.~\ref{fig:BKA4gtd}, $C_1$ is expected to be negative there.
The path labeled $n'$ in Fig.~\ref{fig:BKA4gtd} is now closer to the line 
$\varepsilon_1=0$ than to $\varepsilon_2=0$. In Fig.~\ref{fig:A4f2path} we
find indeed, that a comparison of solutions and approximations~(\ref{eq:G1G2q})
indicates a negative value for $C_1$ to account for a steeper slope 
$(B-B_1-C_1)$. Despite those small deviation, the correlators are described 
well by the asymptotic laws.

The comparison of Figs.~\ref{fig:A4f1path} and~\ref{fig:A4f2path} is summarized
as follows. For each value of $q$ there exists a line with vanishing quadratic
correction $B_2(q)$ for the specified $q$ in the approximation of 
Eq.~(\ref{eq:G1G2q}). On this line, the logarithmic decay is displayed best for
the correlator specified by $q$. Moving to control parameter values above this 
line, $B_2(q)<0$, introduces concave decay of the correlator $\phi_q(t)$ in 
$\ln t$. Going below the line, $B_2(q)>0$, yields a convex decay in the 
correlator $\phi_q(t)$. For increasing the value of $K_q$, the curves specified
by $B_2(q)=0$ rotate clockwise around the $A_4$-singularity.

\begin{figure}[htb]
\includegraphics[width=\columnwidth]{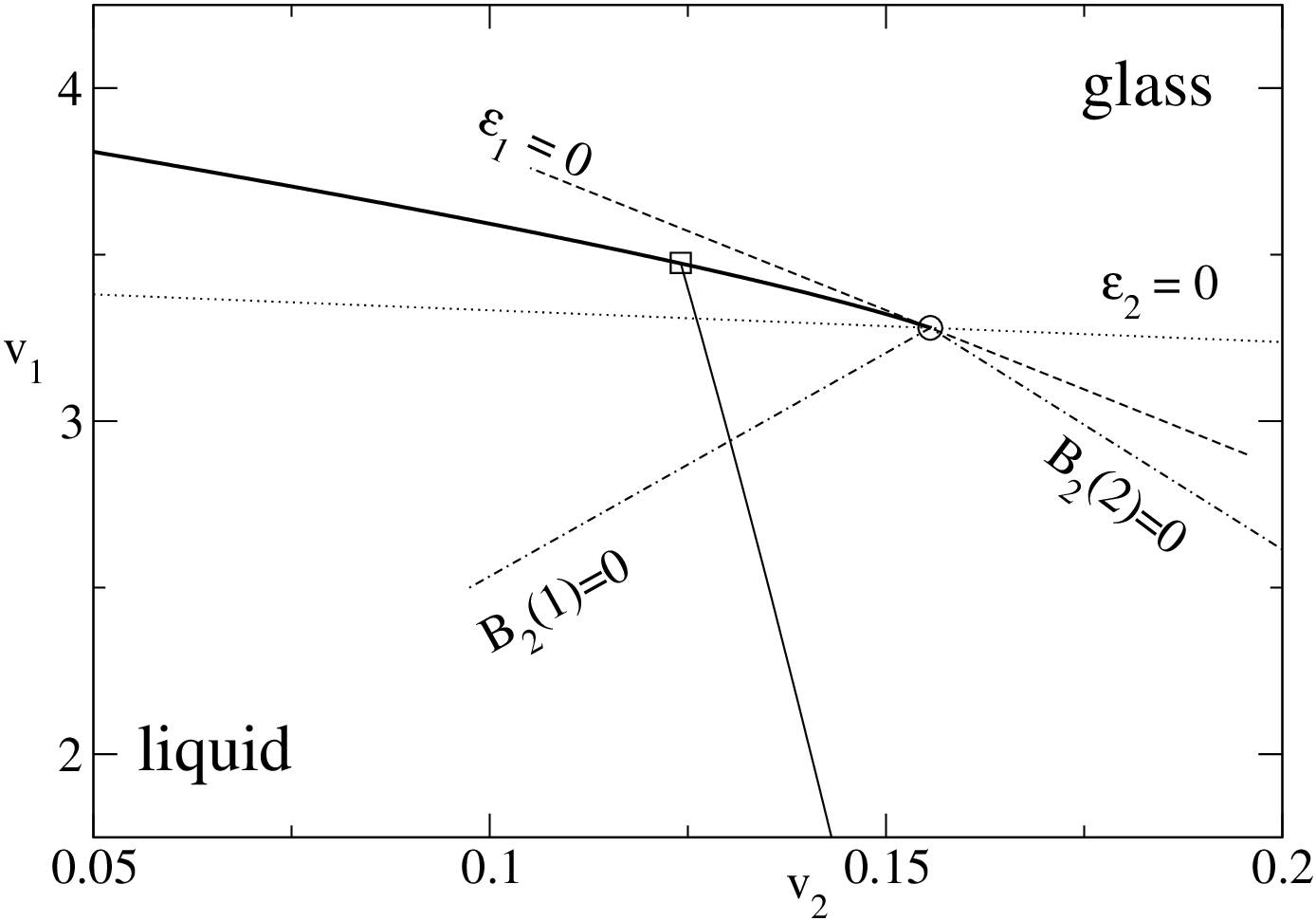}
\caption{\label{fig:BKgtd45}
Glass-transition diagram for the two-component model for $v_3=45$. Notation and
line styles are the same as in Fig.~\ref{fig:BKA4gtd}. The circle marks the 
$A_3$-singularity. Heavy and light full lines denote $A_2$-transition points
with their respective $f_q^c$ higher and lower than the value $f_q^\circ$ at 
the $A_3$-singularity. The crossing point of the two lines is indicated by a 
square.
}
\end{figure}
For control-parameter values $v_3>v_3^*$, the $A_4$-singularity is replaced by
an $A_3$-singularity which terminates a glass-glass-transition line that 
extends further into the arrested region as $v_3$ increases. The lines 
$\varepsilon_1=0$, $\varepsilon_1=0$, $B_2(1)=0$, and  $B_2(2)=0$ for the
$A_3$-singularities emerge from smooth transformations of the ones at the 
$A_4$-singularity. Figure~\ref{fig:BKgtd45} shows a cut through the transition
diagram for $v_3=45$. Different from the situation in Fig.~\ref{fig:BKA4gtd},
the line $B_2(1)=0$ is now well separated from liquid-glass-transition lines
while the line $B_2(2)=0$ is located completely in the arrested regime. The 
dynamics on the line $B_2(1)=0$ for this $A_3$-singularity and the related 
crossing scenario on the same line is discussed in Ref.~\cite{Goetze2002} in
great detail.

\section{Time scale}

It is obvious from Figs.~\ref{fig:A4f1path} and \ref{fig:A4f2path} that the 
time scales $\tau$ for a given state determined by matching, e.g., $\phi_1(t)$
at $f_1^*$ or $f_1^*+\hat{f}_1$ can deviate considerably. The resulting scale
is also different if $\tau$ is fixed for different correlators. However, 
asymptotically close to the higher-order singularity, $\hat{f}_q$ approaches 
zero and also the time scales determined by using different correlators 
converge towards each other as explained earlier \cite{Goetze2002}. Therefore,
only the scale fixed by $\phi_1(\tau)=f_1^*$ shall be considered in the 
following.

\begin{figure}[htb]
\includegraphics[width=\columnwidth]{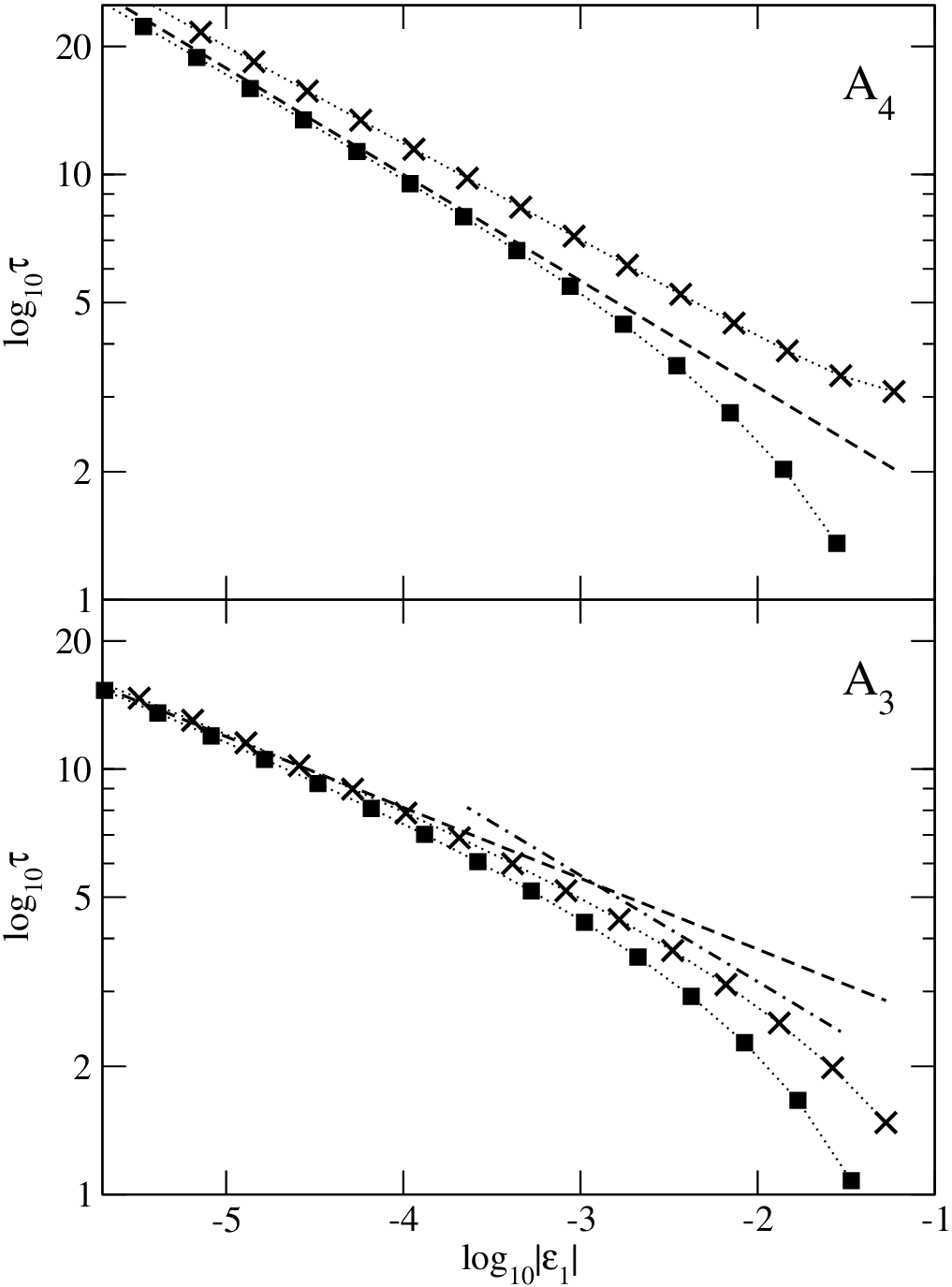}
\caption{\label{fig:BKtau}
Time scale $\tau$ determined by matching $\phi_1(t)$ at $f_1^*$ or $f_1^\circ$
for the paths 
$B_2(1)=0$ (crosses) and $B_2(2)=0$ (squares) from Figs.~\ref{fig:BKA4gtd} and 
\ref{fig:BKgtd45}, respectively. The dotted lines are guides to the eye. The 
corresponding asymptotic laws (dashed lines), $\tau \propto
\exp[1/|\varepsilon_1|^{1/4}]$ for the $A_4$ and $\tau \propto
\exp[1/|\varepsilon_1|^{1/6}]$ for the $A_3$ are fitted to the data for the 
paths where $B_2(1)=0$ around $\tau\approx10^{100}$. The asymptotic law for 
the $A_4$-singularity is redrawn in the lower panel as chain line.
}
\end{figure}
The time scale $\tau$ when considering only the leading order result in 
Eq.~(\ref{eq:G1G2q}) is given by \cite{Goetze1989d} 
\begin{subequations}\label{eq:tau}
\begin{equation}\label{eq:tau:A3}
\log_{10}\tau \propto1/|\varepsilon_1|^{1/6}
\end{equation}
for the $A_3$-singularity and from a similar argument \cite{Flach1992} for 
the $A_4$-singularity as 
\begin{equation}\label{eq:tau:A4}
\log_{10}\tau \propto1/|\varepsilon_1|^{1/4}\,.
\end{equation}
\end{subequations}
Thus, the logarithm of the time scale is the relevant quantity to be discussed 
further on, and $\log_{10}\tau$ is expected to diverge like a power law when
the separation from the singularity vanishes. Figure~\ref{fig:BKtau} shows that
this divergence is indeed stronger for the $A_4$-singularity than for the 
$A_3$-singularity. For $|\varepsilon_1|<10^{-3}$ the time scale $\tau$ is
described qualitatively by the asymptotic laws. The laws in Eq.~(\ref{eq:tau})
were verified quantitatively by extending the analysis to times as large as
$\log_{10}\tau\approx 100$. With one exception, the time scales deviate to 
lower values for $\tau$ for larger separations $|\varepsilon_1|$. Hence, the 
slope of the $\log_{10}\tau$-versus-$\log_{10}(-\varepsilon_1)$ curve is even 
larger in that regime than given by the asymptotic laws in Eq.~(\ref{eq:tau}). 
In the lower panel for $|\varepsilon_1|>10^{-3}$, the time scales for the 
$A_3$-singularity are described better by the law for the close-by 
$A_4$-singularity.

It is seen for the case of the $A_4$-singularity that the proximity of a line 
of liquid-glass transitions (scales $\tau$ indicated by crosses) influences 
also 
the validity of the asymptotic law for $\tau$: Only for extremely large times, 
$\tau$ follows the law~(\ref{eq:tau:A4}). The time scales are also much larger 
on that path as discussed already in connection with Fig.~\ref{fig:A4f1path}. 

\section{Conclusion}
It was shown that the asymptotic expansion in Eq.~(\ref{eq:G1G2q}) provides a 
way to divide the control-parameter space into distinct regions by setting the
dominant correction to the logarithmic decay laws to zero for different 
correlators. If plotted in the form of Figs.\ref{fig:BKA4gtd} and 
\ref{fig:BKgtd45}, the glass-transition diagrams of the two-component model
are easily seen to be topologically equivalent to the ones shown for the 
square-well system in
Refs.~\cite{Sperl2003a,Sperl2003bpre}. For increasing $K_q$ in  
Eq.~(\ref{eq:G1G2q}), the lines of vanishing quadratic corrections rotate 
clockwise around the higher-order singularity in the two-component model, which
mimics the behavior in the microscopic model for increasing wave vector.

Using the asymptotic expansion to describe the solutions of the equations of 
motion~(\ref{eq:BK_eom}), the approximations fit rather accurately already in
a regime that should be accessible to experiments and molecular-dynamics
simulations. However, the presence of other glass-transition singularities in 
the vicinity of higher-order singularities can have drastic influence on the
window in time where the logarithmic laws are applicable, cf. 
Figs.~\ref{fig:A4f1path} and \ref{fig:A4f2path}. The validity of 
Eq.~(\ref{eq:G1G2q}) for $\phi_q(t)$ around $f_q^*$ can be bound either from 
above at earlier times, cf. Fig.~\ref{fig:A4f1path}, or from below to later 
times, Fig.~\ref{fig:A4f2path}, depending on the specific location of the 
chosen path in the transition diagram.

The time scale $\tau$ asymptotically follows the Vogel-Fulcher-type 
law of Eq.~(\ref{eq:tau}). That the relaxation processes close to higher-order
glass-transition singularities are extremely slow is manifested by the fact, 
that instead of $\tau$ itself, the logarithm of the time scale is the relevant
quantity that diverges in some power law. The different behavior of $\tau$ for
$A_4$- and $A_3$-singularities is clearly seen in the numerical solution of
the MCT equations, cf. Fig.~\ref{fig:BKtau}. However, these asymptotic laws 
only show up for extremely long times and small distances from the singularity.
In addition, other close-by glass-transition singularities can significantly
modify the validity of the asymptotic law especially for further separation
from the higher-order singularity under consideration.

\begin{theacknowledgments}
Fruitful discussion with W.~G\"otze is gratefully acknowledged. This work was
supported by the Deutsche Forschungsgemeinschaft Grant No.~Go154/13-2.

\end{theacknowledgments}

%\bibliographystyle{aipproc}
%\bibliography{lit,add}

\end{document}